# Gamma Ray Flashes Produced by Lightning Observed at Ground Level by TETRA-II

D. J. Pleshinger[1], S. T. Alnussirat[1], J. Arias[2], S. Bai[3], Y. Banadaki[3], M. L. Cherry[1], J. H. Hoffman[1], E. Khosravi[3], M. D. Legault[4], R. Rodriguez[5], D. Smith[1], D. Smith[6], E. del Toro[7], J. C. Trepanier[6], and A. Sunda-Meya[8]

[1]Department of Physics & Astronomy, Louisiana State University, Baton Rouge, LA, USA, [2]Centro Nacional de Metrologiá de Panamá, Panama City, Panama, [3]Department of Computer Science, Southern University, Baton Rouge, LA, USA, [4]Department of Physics, University of Puerto Rico at Bayamón, Bayamón, Puerto Rico, [5]Programa de Ciencias Sociales, University of Puerto Rico at Utuado, Utuado, Puerto Rico, [6]Department of Geography and Anthropology, Louisiana State University, Baton Rouge, LA, USA, [7]University of Puerto Rico at Utuado, Utuado, Puerto Rico, [8]Department of Physics of Louisiana, Xavier University, New Orleans, LA, USA

**Abstract** In its first 2 years of operation, the ground-based Terrestrial gamma ray flash and Energetic Thunderstorm Rooftop Array (TETRA)-II array of gamma ray detectors has recorded 22 bursts of gamma rays of millisecond-scale duration associated with lightning. In this study, we present the TETRA-II observations detected at the three TETRA-II ground-level sites in Louisiana, Puerto Rico, and Panama together with the simultaneous radio frequency signals from the lightning data sets VAISALA Global Lightning Dataset, VAISALA National Lightning Detection Network, Earth Networks Total Lightning Network, and World Wide Lightning Location Network. The relative timing between the gamma ray events and the lightning activity is a key parameter for understanding the production mechanism(s) of the bursts. The gamma ray time profiles and their correlation with radio sferics suggest that the gamma ray events are initiated by lightning leader activity and are produced near the last stage of lightning leader channel development prior to the lightning return stroke.

## 1. Introduction

Lightning provides one of the most powerful natural high-energy charged particle accelerators available on Earth. Satellite instruments have detected terrestrial gamma ray flashes (TGFs)—intense submillisecond bursts of bremsstrahlung photons at energies in excess of tens of megaelectron volts emitted by electrons and positrons accelerated by the electric fields associated with thunderstorms. TGFs were initially detected by the Burst and Transient Source Experiment onboard the Compton Gamma-Ray Observatory (Fishman et al., 1994), followed by the Reuven Ramaty High Energy Solar Spectroscopic Imager (Smith et al., 2003), the BeppoSAX satellite (Ursi et al., 2017), and the Relativistic ELECtrons experiment (Panasyuk et al., 2016) and are currently being observed in space by the Gamma ray Burst Monitor (Roberts et al., 2018), the AstroRivelatore Gamma a Immagini Leggero (Marisaldi et al., 2010), the Large Area Telescope (Grove et al., 2012), and the Atmosphere-Space Interactions Monitor (Ostgaard et al., 2018). These observations, associated mainly with intracloud lightning that produces upward moving negative charge, have been correlated with regions of intense lightning and source regions at the altitudes of thunderstorm tops, typically 10–15 km above ground level. In the atmosphere, a single event has been seen at aircraft altitude by ADELE (Smith et al., 2011). Bursts of X-rays and gamma rays associated with lightning have also been reported at ground level by different experiments (Dwyer et al., 2004, 2005; Enoto et al., 2017; Hare et al., 2016; Mallick et al., 2012; Ringuette et al., 2013; Tran et al., 2015; Wada et al., 2019), and related events have been detected by high-energy cosmic-ray detectors (Abbasi et al., 2018; Chilingarian et al., 2010).

Two main mechanisms are discussed as the source of TGFs. The first mechanism involves acceleration by Relativistic Runaway Electron Avalanche (RREA) along with a feedback process (Dwyer, 2003, 2012; Gurevich et al., 1992; Roussel-Dupré & Gurevich, 1996). In this case, relativistic seed electrons are accelerated by the large-scale ambient electric field between opposite polarity charge layers in a thundercloud. The accelerated electrons will knock off secondary electrons by electron impact ionization, which will themselves be accelerated and produce avalanches of electrons and positrons. These electrons will produce photons via bremsstrahlung, and the photons will generate electron-positron pairs. In the feedback process, positrons or backscattered photons moving in the opposite direction to the electrons will create new avalanches of electrons (Dwyer, 2012). The combination of the feedback process with RREA can produce photon fluxes and spectra similar to those seen in satellite-detected TGFs. In this scenario, TGFs may be emitted independent of lightning activity.

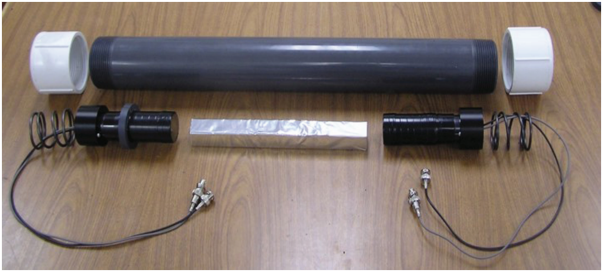

**Figure 1.** The polyvinyl chloride assembly containing a Bismuth Germanate crystal and two photomultipliers.

The second TGF production mechanism involves bremsstrahlung by electrons accelerated in the lightning leader channel (Celestin & Pasko, 2011; Xu et al., 2012). Here ambient electrons can be accelerated to relativistic energies by the very strong small-scale and inhomogeneous electric fields at the tip of the lightning leader channel. In this scenario, TGFs should be emitted prior to the strong optical emission produced by the return stroke. It is also possible that a combination of both of these mechanisms can explain the observed TGF events.

In this paper, we report the detection by the upgraded TGF and Energetic Thunderstorm Rooftop Array (TETRA)-II of 22 X-ray/gamma ray bursts observed at ground level, simultaneously with radio frequency (RF) emission. In section 2, we describe the TETRA-II instrumentation. In section 3, we describe the observations and the correlations with the lightning data sets Earth Networks Total Lightning Network (ENTLN), VAISALA Global Lightning Dataset (GLD360), VAISALA National Lightning Detection Network (NLDN), and World Wide Lightning Location Network (WWLLN) lightning ground detection networks. Discussion and conclusions are presented in sections 4 and 5.

## 2. TETRA-II

TETRA-II consists of three separate arrays of Bismuth Germanate (BGO) scintillators located at Louisiana State University (LSU) in Baton Rouge, Louisiana; the University of Puerto Rico in Utuado; and at the Centro Nacional de Metrologiá de Panamá (CENAMEP) in Panama City, Panama. Locations were chosen due to their high lightning frequency, available infrastructure, and, in the case of Panama and Puerto Rico, their locations below the Fermi satellite orbit. In Utuado, the detectors are deployed on the roof of Building B at University of Puerto Rico in Utuado (lat-long = 18.25°, −66.72°), approximately 25 km from the Atlantic coast on the northern slopes of the Cordillera Central mountains at an altitude of 180 m. In Baton Rouge, the detectors are mounted on the roof of the LSU Physics & Astronomy building (30.41°, −91.17°) approximately 125 km inland from the Gulf of Mexico at an altitude of 15 m above sea level. The Panama City detectors are mounted on the roof of CENAMEP (9.00°, −79.58°) 60 km from the Atlantic coast and 10 km from the Pacific coast at an altitude of 30 m.

A total of 17 detector boxes (10 in Utuado, 5 at CENAMEP, and 2 at LSU) was deployed, each containing six 25.4 × 2.5 × 2.5 cm BGO scintillators capable of detecting gamma rays over the range 50 keV to 6 MeV. Each individual scintillator is wrapped in Al foil and viewed by two 3.8-cm-diameter Hamamatsu R11102 photomultipliers (PMTs) with green sensitivity matched to the BGO output spectrum. One PMT is mounted on each end of the BGO, held in place by a spring and mounted into a polyvinyl chloride tube for protection, as shown in Figure 1. Energy calibrations were performed using various gamma ray sources, including $Na^{22}$ (511 keV, 1.2 MeV), $Cs^{137}$ (662 keV), and thoriated tungsten welding rods (239 keV up to 2.6 MeV), as well as with background peaks seen at each location. At each site, one BGO has been removed from a single box while keeping the remainder of the power and electronics chain active, thereby allowing one electronics channel to monitor for electronic noise.

Each box is divided into two separate devices, each device with its own National Instruments 6351 PCIe high speed (1 Msample/s) data acquisition card handling three BGO (six PMTs). Each PMT anode output pulse is split: One line feeds a fast discriminator used to determine the pulse timing; the other line feeds an amplifier, shaper, and peak-and-hold circuit which is strobed continuously, with the analog-to-digital convertor (ADC) cycling through the PMTs in the device to measure the peak amplitude of each individual PMT. This allows timing of individual PMT pulses to 50 ns and an analog readout of each PMT every 13 μs. The system triggers on coincident signals from a pair of PMTs viewing a single BGO above a ~200-keV threshold or on a GPS pulse-per-second. Timing is determined from a 20-MHz clock, with the pulse-per-second used for realignment every second. Using the 20-MHz clock, individual PMT anode pulse times are recorded with 50-ns resolution. The 50-ns time stamp is used to identify cases in which a single BGO is triggered more than once during the 13-μs device sampling time; in such cases, event timing is recorded, but these events are not used for the energy measurement. When a coincidence is seen between the two ends of a single BGO, the pulse heights of all PMTs in the device are recorded. The range of the ADC is approximately 50 keV to 6 MeV.

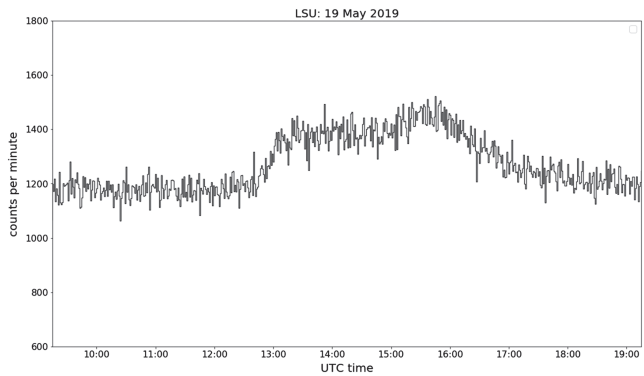

**Figure 2.** Enhancement in Bismuth Germanate event rate measured at Louisiana State University (LSU) on 19 May 2019 at the same time as a thunderstorm.

Figure 2 shows a plot of the count rate per minute in a single device (three scintillators and six PMTs) for a thunderstorm at LSU on 19 May 2019. Prior to the arrival of the thunderstorm at ∼13:00 UTC, the counting rate due to background radioactivity and cosmic rays is ∼1,200/min. (The energy deposition due to a single minimum ionizing cosmic ray passing vertically through a 2.5-cm-thick BGO is typically ∼26 MeV, well above the 6 MeV upper range of the ADC, and so cosmic rays are efficiently identified as saturating events.) The counting rate then increases from ∼1,200/min to ∼1,400/min over the several hours of the storm. Ringuette et al. (2013) have shown a similar time history from TETRA-I and demonstrated the presence of prominent radon lines during the period of increased count rate.

The data analysis searches for event candidates in each device by binning the entire day into 2-ms bins and selecting events exceeding 20 standard deviations above the daily average background in both devices in a single box. It should be noted that this 20 $\sigma$ selection criterion has no relationship to random or Gaussian statistics. The criterion of 20 $\sigma$ above the daily average is simply a convenient means of identifying 2-ms excesses above the background of cosmic rays, background radiation, and increased counting rate on scales of hours due to the overall thunderstorm enhancement seen in Figure 2.

Application of the 20 $\sigma$ criterion is done separately for each device since each has its own typical background rate. If two devices in a single box see a candidate that meets this 20 $\sigma$ criterion, the remaining devices at the location are further examined. The 20 $\sigma$ threshold was determined initially from the previous TETRA-I results (Ringuette et al., 2013) and confirmed for TETRA-II as the point where events associated with nearby lightning begin to stand out above the typical background. Figure 3 shows the number of 2-ms bins observed in each device versus the significance over the background rate for data within 5 s and 8 km of a lightning strike (dark line) and all data (shaded gray). Although a loose cut of 8 km is set for nearby storms, we expect neither photons nor electrons from distances greater than ∼2 km due to atmospheric attenuation. The gray background data are normalized to the same live time (∼36 hr) as the data with nearby lightning during the days events were observed. As shown in the inset, the black (nearby lightning) data begin to exceed the gray (background) data near 20 $\sigma$. When account is taken for the fact that a single event registers in multiple devices (up to a maximum of six devices in four boxes spaced up to a maximum distance of approximately 20 m in the event on 22 October 2018 in Panama), 22 separate events satisfy the 20 $\sigma$ threshold condition.

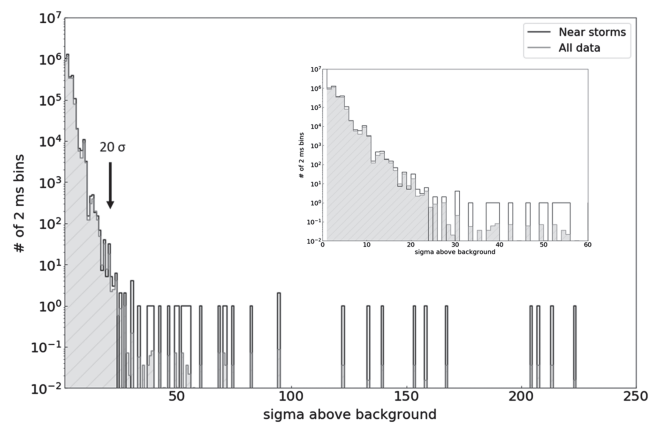

**Figure 3.** Number of 2-ms bins in a single device versus significance above daily background. Black line shows events within 5 s and 8 km of a lightning strike ("near lightning" events); shaded gray shows all data ("background") normalized to the same live time as the "near lightning" sample (∼36 hr). An excess of events associated with lightning is seen for $\sigma \geq 20$. Events are plotted for $\sigma \geq 1$ only. Inset shows an expanded view of the region up to 60 sigma.

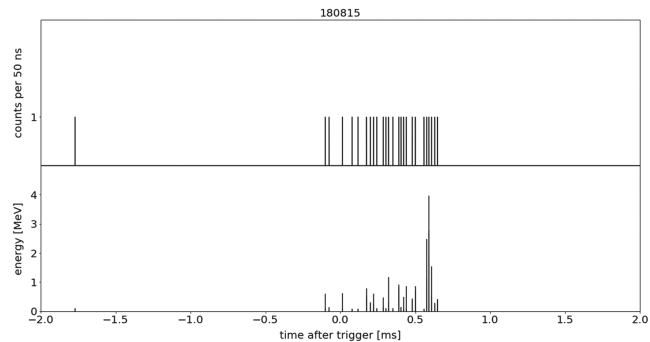

**Figure 4.** (top) Time series for 180815 event seen at Louisiana State University for a single device with three Bismuth Germanate scintillators. (bottom) Energy versus time for same device as time series above.

In parallel with the analysis searching for two devices in a single box above the 20 $\sigma$ threshold, a second analysis is performed looking for a single device with more than 5 counts in a single 1-ms bin. This removes the requirement that both devices must observe the event. If a single device at a location observes a candidate above this threshold, all the remaining devices at the location are checked. This second search has found the same 22 events.

An example of a time history of a 20 $\sigma$ event is shown in Figure 4. Here counts per 50-ns bin are shown for a 4-ms window for one device at LSU for the event on 15 August 2018.

The event candidate times are compared with radio data from the NLDN, the GLD360, the ENTLN, and the WWLLN. NLDN data are used for comparison with LSU, GLD360 for Utuado and Panama, and ENTLN and WWLLN are obtained for all locations.

## 3. Observations

TETRA-II began operation in Fall 2015 with the installation of both boxes at LSU, with the remaining locations becoming operational over the next 18 months. To date, TETRA-II has detected 22 TGF-like events in Louisiana, Panama, and Puerto Rico (Table 1) typically ranging in duration from 0.1 to 2 ms. (Initially, two detector boxes were installed at the Severe Weather Institute—Radar and Lightning Laboratories of the University of Alabama in Huntsville, Alabama. After no events were seen in Huntsville in 1.5 years, the detectors were removed from Huntsville in Fall 2018.) Once an event is found, the time stamp is compared with the lightning catalogs, searching for a coincident NLDN, GLD360, ENTLN, or WWLLN strike nearby within 8 km of the detector location and 5 s of the trigger time. Of the 22 events detected, 19 have a radio sferic detection within 6 km and 1.5 ms of the beginning of the gamma ray event (Table 2). In every case, the radio signal is reported after the start of the gamma ray event, usually close to or at the end of the event.

Of the 12 events TETRA-II detected at LSU, eight were accompanied by an NLDN strike within 1.0 km and 1.0 ms (Table 2). All eight were negative polarity cloud-to-ground (CG) strikes, with peak currents ranging from −22.3 to −111.4 kA. Two events detected at LSU were associated with ENTLN sferics within 0.5 km and 1.14 ms. Both of these were also negative polarity CG strikes. Two LSU events had no associated radio signal.

**Table 1**
*Number of Events and Operating Time for Each Location*

| Location | Live time (years) | Events |
|---|---|---|
| Louisiana | 3 | 12 |
| Puerto Rico | 2.5 | 1 |
| Panama | 1.5 | 9 |
| Alabama | 1.5 | 0 |

**Table 2**
*Terrestrial Gamma Ray Flashes and Energetic Thunderstorm Rooftop Array-II Event Properties*

| Event | Location | Time stamp UTC | Counts | Duration (µs) | $\Delta t$ (µs) | Lightning distance (km) | Lightning source |
|---|---|---|---|---|---|---|---|
| 160427 | LSU | 16:49:25.418 | 19 | 100 | 20 | 1.0 | NLDN |
| 160919 | Utuado | 18:09:33.762 | 183 | 800 | 530 | 2.6 | GLD360 |
| 170307 | LSU | 23:34:30.446 | 169 | 700 | 50 | 0.5 | NLDN |
| 170325a | LSU | 15:47:15.270 | 73 | 500 | 470 | 0.2 | NLDN |
| 170325b | LSU | 16:02:12.737 | 29 | 450 | 460 | 0.4 | NLDN |
| 170325c | LSU | 16:02:12.918 | 61 | 250 | 230 | 0.4 | NLDN |
| 170601 | Panama | 01:15:24.179 | 23 | 850 | 700 | 5.9 | GLD360 |
| 170624a | LSU | 19:34:50.268 | 203 | 1,150 | 1,080 | 0.4 | NLDN |
| 170624b | LSU | 19:34:50.475 | 133 | 400 | 350 | 0.5 | NLDN |
| 170624c | LSU | 19:34:50.364 | 48 | 100 | 10 | 0.5 | NLDN |
| 170707 | LSU | 22:25:51.186 | 113 | 5,950 | * | * | * |
| 170810a | Panama | 14:34:01.703 | 91 | 1,350 | 1,300 | 0.6 | GLD360 |
| 170810b | Panama | 14:34:01.684 | 19 | 350 | * | * | * |
| 171018a | Panama | 17:43:46.565 | 97 | 1,550 | 1,180 | 2.5 | GLD360 |
| 171018b | Panama | 17:45:31.545 | 34 | 900 | 850 | 2.9 | GLD360 |
| 171103 | Panama | 19:34:30.382 | 25 | 350 | 190 | 6.5 | WWLLN |
| 171204 | Panama | 17:54:50.349 | 24 | 1,750 | 880 | 3.5 | GLD360 |
| 180605 | Panama | 11:59:21.008 | 44 | 650 | 590 | 6.8 | WWLLN |
| 180815 | LSU | 22:56:43.222 | 56 | 950 | 890 | 0.5 | ENTLN |
| 180817 | LSU | 13:51:59.767 | 45 | 650 | 1,140 | 0.5 | ENTLN |
| 180915 | LSU | 20:42:56.859 | 15 | 500 | * | * | * |
| 181022 | Panama | 21:54:00.386 | 89 | 1,300 | 1,010 | 1.1 | WWLLN |

*Note.* Duration is defined as time from the first to the last photon. $\Delta t$ is defined as the time difference between the radio sferic and $t_0$. Lightning distance is given as the distance to the radio position, and $I_P$ is the peak current determined from the radio measurement. Counts in events are included. Asterisk (*) denotes no reported values. NLDN = VAISALA National Lightning Detection Network; GLD360 = VAISALA Global Lightning Dataset; ENTLN = Earth Networks Total Lightning Network; WWLLN = World Wide Lightning Location Network.

Figure 5 shows one of three events from the detectors at LSU on 25 March 2017. Time zero is defined as the first 50 µs bin with ≥2 counts. The radio signal is labeled by the plus sign, with the distance to the radio event given on the right-hand axis. The burst was seen to build up over its ~50- µs duration and then promptly cut off after its peak, 50 µs before the NLDN radio signal.

For both Utuado and Panama City, GLD360 data were used to study the correlation between RF sferic and TETRA-II gamma ray events. Of the nine events seen in Panama, five have a GLD360 strike within 2 ms from the beginning of the event and within 6.5 km of the detectors, four associated with negative polarity lightning, and one with a positive polarity stroke. Three of the Panama events were associated with a WWLLN strike within 0.6 ms of the beginning of the gamma ray trigger and within 7 km.

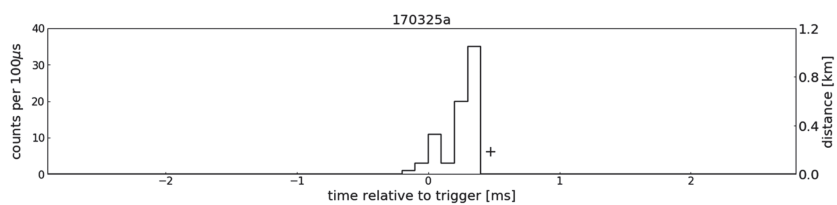

**Figure 5.** Time series of photons observed over a 6-ms window binned in 100-µs bins for the Louisiana State University event at 15:47:15 UTC on 25 March 2017, summed over the 11 Bismuth Germanate in the two boxes.

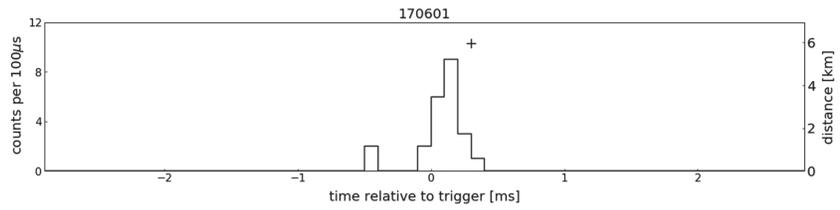

**Figure 6.** Time series over a 6-ms window binned in 100-μs bins for the Panama event at 01:15:24 UTC on 1 June 2017, summed over all Bismuth Germanate in three boxes.

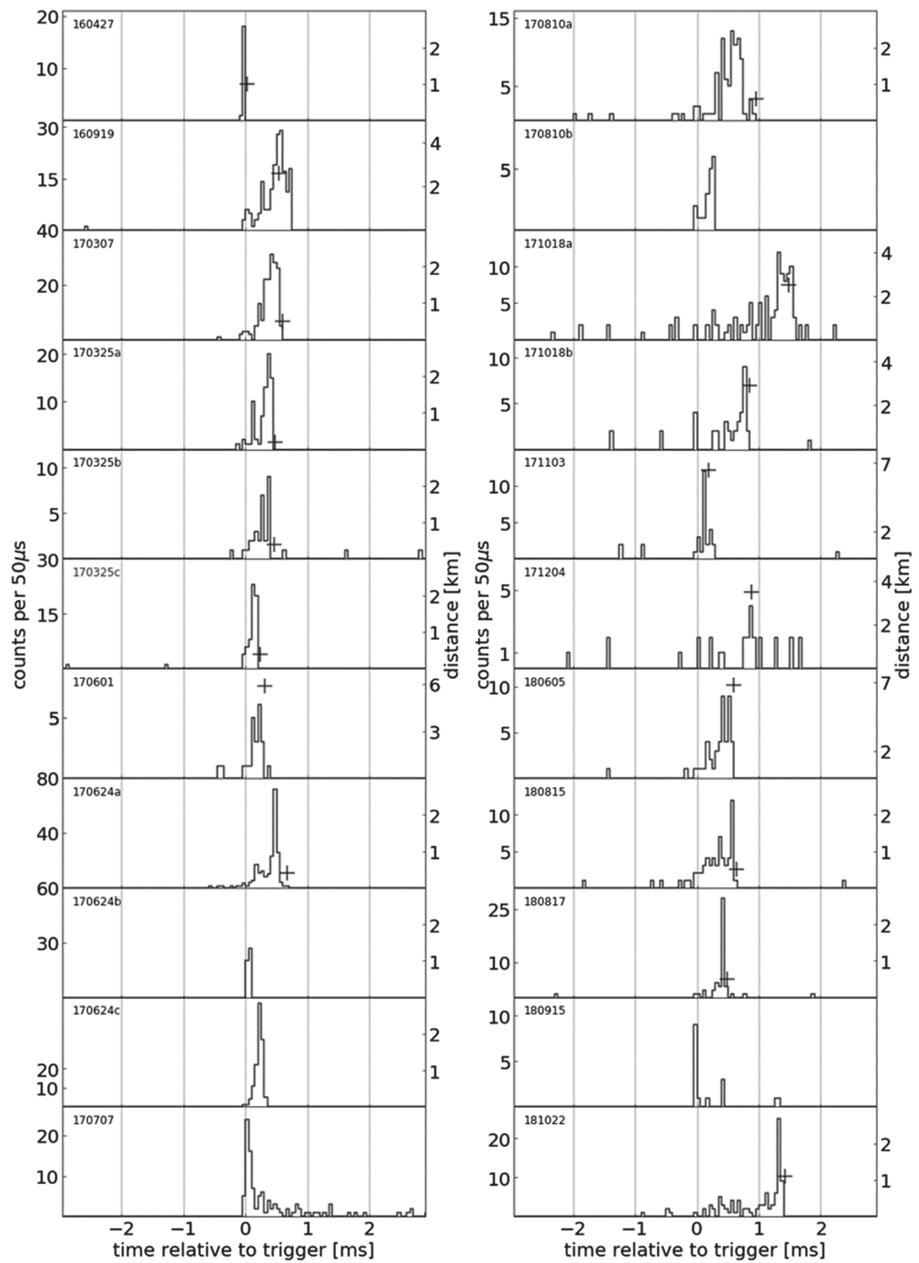

**Figure 7.** Counts per 50 μs for 22 Terrestrial gamma ray flash and Energetic Thunderstorm Rooftop Array-II events. Time of associated radio signal is shown as +, with distance to the reported radio event shown on the right-hand scale.

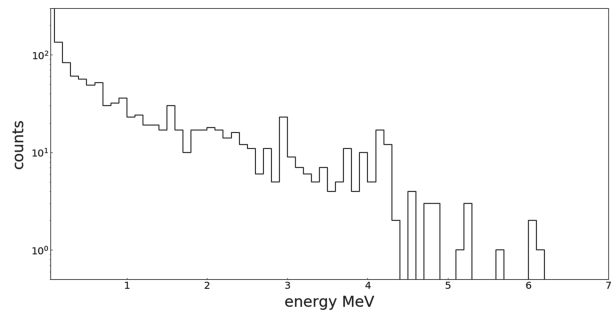

**Figure 8.** Summed energy spectrum of photons from the 22 Terrestrial gamma ray flash and Energetic Thunderstorm Rooftop Array-II events.

Figure 6 shows the event at CENAMEP in Panama on 1 June 2017. This event is roughly symmetrical in shape over the ∼500-μs duration of the main peak. In this case, the radio signal observed from GLD360 is reported ∼100 μs after the peak of the burst.

Figure 7 shows the time series for all 22 events over 6-ms windows in 50-μs time bins. Event durations range from 100 μs to 6 ms. The average number of photons detected is 70, with average burst duration 980 μs.

An average energy spectrum summed over 22 events is presented in Figure 8. Energies were determined by converting the recorded ADC channel values using the calibration data from known X-ray sources and cosmic ray muons. The ADC was observed to behave linearly up to 10 MeV. Energy deposits up to 6 MeV are seen. Further analysis of the energy spectra will be presented in a future paper.

In summary, 19 of the 22 TGF-like TETRA-II events had a lightning strike reported from at least one of NLDN, GLD360, ENTLN, and WWLLN. In every case, the lightning strike occurred between 13 μs and 1.3 ms after the beginning of the event, typically at the end of the gamma ray event. All of the lightning strikes were within 7 km, with nine reported less than 500 m away. Out of the 16 events where the lightning polarity is known, 15 are negative and 1 was positive, and in all 10 cases where the type is known, the type is CG. Half the events show the characteristic behavior illustrated in Figure 5, with the signal increasing over the duration of the event and then abruptly ending.

## 4. Discussion

For a typical negative CG lightning stroke, the stepped leader process can take up to 35 ms, with steps 50 m in length and up to 50 μs apart (Rakov & Uman, 2003). For all the observed TETRA-II events with an associated lightning strike, Table 2 shows a positive time offset, $\Delta t$, between $t_0$ of the gamma ray signal and the radio sferic time with $\Delta t \leq 2$ ms in every case, suggesting that these events are produced not only by the lightning leader process but also during the last stages of the channel development.

Eleven events show a similar structure to that in Figure 5 with the burst building up and then terminating shortly after its peak, with the radio signal following. This structure suggests bursts emitted by individual lightning leader steps as they move downward toward the detectors. As each step approaches, the number of photons detected increases until the gamma ray production is terminated by the stroke hitting the ground, followed by the return stroke accompanied by the RF emission.

One unique event is shown in Figure 9. This event, detected on 7 July 2017 at LSU, shows a sharp spike at the beginning of the event and then a long tail lasting up to 6 ms. Assuming the bursts observed by TETRA-II are beamed with half-angle $\Delta\Theta$, as is seen from space, where $\Delta\Theta \sim 30°$ (Connaughton et al., 2010), then this event can be understood as an event directed away from the detectors but at an angle $< \Delta\Theta$ with respect to TETRA-II; in this case, successive leaders are farther from the detector, and the number of events detected decreases with increasing time and distance. No nearby sferic was reported in association with this event, although lightning was present at a distance of slightly more than 8 km.

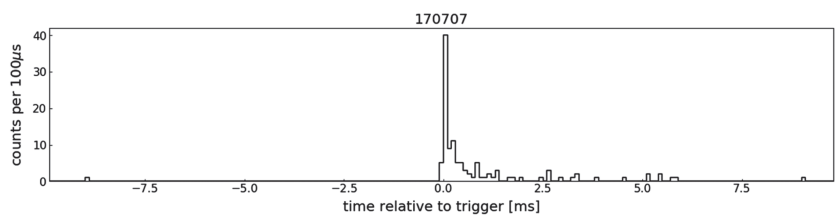

**Figure 9.** Time series over a 20-ms window binned in 100-μs bins for the Louisiana State University event at 22:24:51 UTC on 7 July 2017.

X-ray emission associated with lightning leaders has been reported by others (Dwyer et al., 2004, 2005, 2012; Mallick et al., 2012; Tran et al., 2015), with similarities as well as distinct differences compared to the events presented here. Dwyer (2012) observed, along with a TGF after a lightning return stroke, a collection of X-ray hits that were abruptly cut off at the moment of the lightning return stroke, similar to the observation of Dwyer et al. (2005). Mallick et al. (2012) reported multiple lightning flashes with similar X-ray emission associated with lightning strikes, typically with energies below 1 MeV and durations less than 100 μs. The largest X-ray burst observed by Mallick et al. (2012) lasted ∼200 μs with measured energies up to ∼5 MeV, somewhat shorter in duration than the average TETRA-II event but similar in energy to the events in the TETRA-II sample. One distinct difference is that the large burst reported by Mallick occurred in association with the third stroke of the lightning flash, ∼50 ms after the initial return stroke. The TETRA-II lightning associations are seen with a single reported stroke by the various commercial lightning networks, indicating they are produced from the initial stepped leader process before the first return stroke.

The TETRA-II events in Table 2 have durations ∼100 μs up to ∼1.8 ms (not including the Unusual Event 170707 with a tail extending to ∼6 ms). As pointed out by Dwyer (2012) and Tran et al. (2015), the typical duration of bursts observed from space is ∼100 μs (Briggs et al., 2010; Fishman et al., 2011). Events reported from the Lightning Observatory in Gainesville (Mallick et al., 2012, Figures 5–7; Tran et al., 2015, Figure 1) and the International Center for Lightning Research and Testing at Camp Blanding (ICLRT; Dwyer et al., 2005, Figure 1; Dwyer et al., 2004, Figure 1) have started ∼80–800 μs prior to and terminating at the time of the return stroke. Dwyer et al. (2004) report a burst of X-rays beginning ∼60 μs prior to the return stroke for a triggered lightning event at the ICLRT. On average, these events are somewhat shorter in duration than but not inconsistent with the TETRA-II sample of events. Likewise, the reported energies of the Lightning Observatory in Gainesville and ICLRT events range up to ∼10 MeV, similar to TETRA-II. The upward stepping time structure in Figure 5 and the long-duration events (e.g., Figure 9) have not been previously reported.

A number of authors (e.g., Dwyer, 2012; Tran et al., 2015) have drawn a distinction between X-ray/gamma ray flashes associated with lightning leaders versus TGFs. The distinction is significant if the very strong electric fields at the lightning leader tips are able to accelerate bulk-free electrons to high energies in a "cold runaway" process, while in the case of TGFs, the somewhat weaker but large-scale fields in thunderclouds accelerate higher-energy seed electrons via RREA avalanches. The durations of the TETRA-II bursts are typically longer than the events observed from space, but this may be a result of the propagation through the dense atmosphere near ground level compared to the more tenuous atmosphere at high altitudes traversed by upward moving TGFs.

Tran et al. (2015) present an operational definition of TGFs as events with the following characteristics: (a) no sign of pileup of lower-energy X-rays masquerading as higher-energy gamma rays, (b) event durations less than 1 ms, and (c) energy of individual photons exceeding 1 MeV. As discussed above, the durations of the TETRA-II events are indeed somewhat longer than the typical TGF events observed from space. This may be a consequence of scattering in the dense atmosphere near ground level. As shown in Figure 8, the energies measured by TETRA-II extend to ∼6 MeV. A detailed analysis of the energy spectrum and event durations of the TETRA-II events will be presented in a future paper.

Event times are recorded with 50-ns resolution, with no indication of pileup. Events occurring within 13 μs of the previous event in the same BGO are included in the time histories, but energies are not recorded. In addition, in 2018, a 5-mm-thick sheet of tin (Sn) was placed above one of the LSU detectors. This is a sufficient thickness so that ∼52% of incident 200-keV X-rays undergo a photoelectric interaction. The ratio

$$R = \frac{\#\ \text{of events} \leq 200\,\text{keV}}{\#\ \text{of events} \geq 200\,\text{keV}} \qquad (1)$$

is a measure of pileup: For the unshielded detector, $R_{unshielded}$ measures the hardness ratio of the incident photons. For the shielded detector, $R_{shielded}$ measures the attenuation of soft photons compared to hard photons. Since 2018, a preliminary measurement gives $R_{unshielded} = 1.5 \pm 0.3$ and $R_{shielded} = 1.0 \pm 0.2$, consistent with no pileup effects.

There have been several studies of the association between TGFs and radio sferics. Roberts et al. (2018) have found an association rate of ∼30% between Gamma ray Burst Monitor TGFs and WWLLN sferics, with most of the TGFs within 200 μs of the reported WWLLN time. Given this close association of at least some TGFs with the lightning leaders and given that the TETRA-II events seem to satisfy the Tran et al. (2015) TGF requirements, we see no evidence that the leader-associated ground-level TETRA-II events are distinctly different from TGFs. That is not to say that all TGFs are associated with lightning leaders, but it does indicate that acceleration in the high field regions of CG lightning leader tips produces downward events very similar to TGFs resulting from upward intracloud lightning observed from space.

## 5. Conclusions

In the original version of the experiment, TETRA-I observed over 30 bursts at LSU with an array of sodium iodide (NaI) detectors (Ringuette et al., 2013). The TETRA-II results presented here show a similar pattern of events associated with nearby negative polarity lightning, with the expected increase in photons detected due to the higher density BGO scintillators. Twenty-two TGF-like events have been observed in the first 2 years of TETRA-II observations. The average number of photons detected over the energy range 200 keV to 8 MeV is 70, with average burst duration 980 μs. In 19 of the bursts, a lighting strike (at distances ranging from 0.2 to 6.5 km) occurred between 13 μs and 1.3 ms after the beginning of the event, supporting the argument that these events were produced during the later stages of the lightning step leader process. A stepping structure was seen as events built up before abruptly ending at the time of an associated lighting flash, suggesting that the gamma rays were typically produced by individual downward moving leaders approaching the detectors. Future studies will look at the details of the energy spectra and time histories and associations with the thunderstorm characteristics. The events observed so far show a broad variation of time structures and frequency at the three TETRA-II locations, indicating the need for continued monitoring and more complete cataloging of these events.

**Acknowledgments**

We gratefully acknowledge funding from NSF (Office of Atmospheric and Geospace Sciences), NASA EPSCoR, the Louisiana Board of Regents, and Louisiana Space Consortium for the development and operation of TETRA-II and the very helpful support of Milton Riutort and the administration at the University of Puerto Rico-Utuado and Franklin Hurley at CENAMEP. Lightning data were provided by Vaisala (https://www.vaisala.com), Earth Networks (https://www.earthnetworks.com/), and WWLLN (http://wwlln.net/). The authors would like to thank N. Cannady for stimulating discussions, and S. B. Ellison, D. Granger, C. Adams, E. Brooks, K. Clear, J. Garriz, R. Pace, and N. Zimmer for their participation in various aspects of the experiment. TETRA-II data are available online (https://tetra.phys.lsu.edu).